\documentclass[aps,prl,twocolumn]{revtex4}

\usepackage{epsfig}
\usepackage{axodraw}
\usepackage{graphicx}

\newcommand{\la}{{\lambda}}

\newcommand{\be}{\begin{equation}}
\newcommand{\ee}{\end{equation}}
\newcommand{\beq}{\begin{equation}}
\newcommand{\eeq}{\end{equation}}
\newcommand{\bea}{\begin{eqnarray}}
\newcommand{\eea}{\end{eqnarray}}
\newcommand{\br}{\begin{eqnarray}}
\newcommand{\er}{\end{eqnarray}}
\newcommand{\ba}{\begin{array}}
\newcommand{\ea}{\end{array}}
\newcommand{\bi}{\begin{itemize}}
\newcommand{\ei}{\end{itemize}}
\newcommand{\bn}{\begin{enumerate}}
\newcommand{\en}{\end{enumerate}}
\newcommand{\bc}{\begin{center}}
\newcommand{\ec}{\end{center}}

\newcommand{\bT}{\bar{T}}
\def\bY{{\bf Y}}
\def\bA{{\bf A}}
\def\bB{{\bf B}}

\def\bU{{\bf U}}
\def\bV{{\bf V}}
\def\bm{{\bf m}}
\def\bM{{\bf M}}

\def\tl{{\tilde{L}}}
\def\tm{{\tilde{m}}}

\def\tq{{\tilde{Q}}}

\def\unity{{\hbox{1\kern-.8mm l}}}

\newcommand{\gsim}{\lower.7ex\hbox{$\;\stackrel{\textstyle>}{\sim}\;$}}
\newcommand{\lsim}{\lower.7ex\hbox{$\;\stackrel{\textstyle<}{\sim}\;$}}

\newcommand{\bfD}{{\Delta}}

\begin{document}

\title{Testing Neutrino Mass Models}


\author{Eung Jin Chun}
\affiliation{Korea Institute for Advanced Study, 207-43
Cheongryangri,
  Dongdaemun, Seoul 130-722, Korea }

\begin{abstract}
The existence of the neutrino masses and mixing would be an
important window into the nature of physics beyond the Standard
Model, which will be searched for in the forthcoming experiments
such as LHC. In this talk, we discuss some examples of neutrino
mass models which are testable through the observation of lepton
flavor violating processes and/or electric dipole moments
correlated with the neutrino mass structure.
\end{abstract}

\maketitle


\section{Introduction and summary}

While the Standard Model is firmly established by various
accelerator experiments in the past, there are also a number of
theoretical and experimental reasons that it is a low energy
effective theory of a more fundamental theory.  The discovery of
neutrino masses and mixing in astrophysical and reactor neutrino
experiments would be one of the most important experimental
evidences for physics beyond the Standard Model.  A large number
of new physics models have been suggested to explain the origin of
the neutrino mass. Among them,   low-energy models at TeV scale
are of the special interest as they can be directly tested in the
forthcoming colliders; LHC or LC. Some high-energy models,
motivated by superstring or grand unification theories, can also
lead to observable predictions providing an indirect test of the
models.

In this talk, we present three examples of such testable neutrino
mass models in anticipation of discovering new physics in the
future collider or low-energy experiments:

1. Higgs  Triplet  model at TeV scale where the neutrino mass
structure can be probed by observing doubly charged higgs bosons
decaying to two charged leptons \cite{1}.

2.  Supersymmetric standard model with R-parity violation where
the lightest supesymmetric particle (LSP) decay encodes the
information of the neutrino masses and mixing \cite{2}.

3.  Supersymmetric  triplet seesaw model at high-energy scale
which may provide testable correlations among lepton flavor
violating (LFV) processes and electric dipole moments (EDMs)
reflecting the neutrino flavor structure \cite{3}.

 Let us now discuss  the main features of each model
 which can be searched for in the future experiments to discover
 the new physics beyond the Standard Model.

\section{Triplet Seesaw Model at TeV scale}

The first example  is  the {\it Higgs triplet model} in which a
triplet scalar field $T=(T^{++},T^+,T^0)$ with the mass $M $ is
introduced to have the following renormalizable couplings;
\begin{equation}
{\cal L}_T = {1\over\sqrt{2}}[ f_{ij} L_i L_j T +
 \mu\Phi\Phi T + h.c.] -M^2 |T|^2\,,
\end{equation}
where $L_i=(\nu_i,l_i)_L$ is the left-handed lepton doublet and
$\Phi=(\phi^0,\phi^-)$ is the standard model Higgs doublet.  Due
to the ``$\mu$'' term in the above equation, the neutral component
$T^0$ of the triplet gets the vacuum expectation value (VEV), $v_T
= \mu v_\Phi^2/2M^2$ where $v_\Phi= \langle \phi^0 \rangle = 246$
GeV. This leads to the neutrino mass matrix,
\begin{equation}
M^\nu_{ij}=  f_{ij}  v_T \,.
\end{equation}
The above relation shows that the nine independent parameters
contained in $f$ are  in {\it one-to-one} correspondence with the
low-energy neutrino parameters. As a consequence unambigous
predictions on the low-energy LFV phenomena can be derived in the
triplet seesaw model.

We look for the possibility of the light triplet Higgs bosons,
namely $M \sim $ TeV, so that observations of various lepton
flavor violating processes can provide a probe for the neutrino
masses and mixing through the relation (2), and thus  a direct
test of the model. In this ``low-energy triplet Higgs model'', the
small parameters $f$ and $\xi \equiv v_T/v_\Phi$ are required;
\begin{equation}
f_{ij} \xi  \sim 10^{-12}
\end{equation}
for $M^\nu_{ij}\sim 0.3$ eV. For sizable couplings $f$, the
exchange of the triplet Higgs can induce the lepton flavor
violating processes like $\mu\to e \gamma$.  In Table I, the
bounds on the couplings $f$ are summarized.

Some of striking collider signals in the triplet Higgs model comes
from the decays of a doubly charged Higgs boson, such as $T^{--}
\to l_i l_j, W^-W^-$, which have been studied extensively in the
literature. We are interested in the situation that the decays
$T^{--} \to l_i l_j$ are sizable so that the neutrino mass
structure can be tested in colliders.   Depending on the masses of
the triplet components, the fast decay process like $T^{--} \to
T^- W^{(*)-}$ through gauge interactions can happen to
over-dominate any other processes of our interest. The mass
splitting among the triplet components arises upon the electroweak
symmetry breaking and thus is of the order $M_W$. The most general
scalar potential for a doublet and a triplet Higgs boson is
\begin{eqnarray}
V&=& m^2 (\Phi^\dagger\Phi) + \lambda_1 (\Phi^\dagger\Phi)^2
  + M^2 \mbox{Tr}(\bfD^\dagger\bfD) \nonumber \\
&&  + \lambda_2 [\mbox{Tr}(\bfD^\dagger\bfD)]^2
  + \lambda_3 \mbox{Det}(\bfD^\dagger\bfD) \nonumber\\
&&  + \lambda_4 (\Phi^\dagger\Phi)\mbox{Tr}(\bfD^\dagger\bfD)
  + \lambda_5 (\Phi^\dagger \tau_i\Phi)\mbox{Tr}(\bfD^\dagger \tau_i \bfD)
\nonumber\\
&&  + {1\over\sqrt{2}}\mu  (\Phi^Ti\tau_2 \bfD \Phi)  +h.c.,
\end{eqnarray}
where $\bfD$ is the 2$\times$2 matrix representation of the
triplet $T$.   In this model, the mass eigenstates consist of
$T^{++}$, $H^+$, $H^0$, $A^0$ and $h^0$. Under the condition that
$|\xi| \ll 1$, the first five states are mainly from the triplet
sector and the last from the doublet sector.

\begin{widetext}
\begin{center}
\begin{tabular}{|c|c|c|c|}
\hline Mode & ~Current limit ~
     & ~Future sensitivity ~
     &~ Bound on the couplings~ \\
\hline $\mu\to e \gamma$ & $1.2\times10^{-11}$ & $\sim10^{-14}$
     & $(f f^\dagger)_{12} < 1.2\times10^{-4}\, x_T$ \\
$\tau\to e \gamma$ & $2.7\times10^{-6}$ & $\sim10^{-8}$
     & $(ff^\dagger)_{13} < 1.3\times10^{-1}\, x_T$ \\
$\tau\to \mu \gamma$ & $0.6\times10^{-6}$ & $\sim10^{-8}$
     & $(ff^\dagger)_{23} < 6.1\times10^{-2}\, x_T$ \\
$\mu\to \bar{e}ee$ & $1.0\times10^{-12}$ & $\sim10^{-15}$
     & $f_{11}f_{12} < 9.3\times10^{-7}\, x_T$ \\
$\tau\to \bar{e}ee$ & $2.7\times10^{-7}$ & $\sim10^{-8}$
     & $f_{11}f_{13} < 1.1\times10^{-3}\, x_T$ \\
$\tau\to \bar{e}e\mu$ & $2.4\times10^{-7}$ & $\sim10^{-8}$
     & $f_{12}f_{13} < 1.5\times10^{-3}\, x_T$ \\
$\tau\to \bar{e}\mu\mu$ & $3.2\times10^{-7}$ & $\sim10^{-8}$
     & $f_{22}f_{13} < 1.2\times10^{-3}\, x_T$ \\
$\tau\to \bar{\mu}ee$ & $2.8\times10^{-7}$ & $\sim10^{-8}$
     & $f_{11}f_{23} < 1.2\times10^{-3}\, x_T$ \\
$\tau\to \bar{\mu}e\mu$ & $3.1\times10^{-7}$ & $\sim10^{-8}$
     & $f_{12}f_{23} < 1.7\times10^{-3}\, x_T$ \\
~$\tau\to \bar{\mu}\mu\mu$~ & $3.8\times10^{-7}$ & $\sim10^{-8}$
     & $f_{22}f_{23} < 1.4\times10^{-3}\, x_T$ \\
\hline
\end{tabular}

\vspace{1ex}
 {TABLE I: The experimental limits on the branching
ratios of various modes and the\\  corresponding upper bounds on
the product of couplings taking $x_T = (M_{T}/200 {\rm GeV})^2$.}
\end{center}
\end{widetext}

When $\lambda_5>0$, we have $M_{T^{\pm\pm}} < M_{H^\pm} <
M_{H^0,A^0}$, so that the doubly charged Higgs boson $T^{--}$ can
only decay to $l_i l_j$ or $W^- W^-$ through the following
interactions;
\begin{equation}
 {\cal L} = {1\over\sqrt{2}} \left[ f_{ij}\, \bar{l^c}_i P_L l_j
  +  g\xi M_W\, W^- W^-  \right] T^{++} + h.c.
\end{equation}
The corresponding decay rates are
\begin{eqnarray}
\Gamma(T^{--}\to l_i l_j) &=& S {f_{ij}^2 \over 16 \pi}
     M_{T^{\pm\pm}} \nonumber \\
\Gamma(T^{--}\to W W) &=& {\alpha_2 \xi^2 \over32}
   {M_{T^{\pm\pm}}^3\over M_W^2}
            (1-4r_W+12r^2_W) \nonumber\\
 &&(1-4r_W)^{1/2}
\end{eqnarray}
where $S=2\,(1)$ for $i\neq j\, (i=j)$ and
$r_W=M_W^2/M_{T^{\pm\pm}}^2$.  In this case, the heavier states
$H^+$, $H^0$ and $A^0$ can have the decay modes; $H^0,A^0 \to H^+
W^{(*)-}$ and $H^+\to T^{++} W^{(*)-}$ leading to the production
of $T^{\pm\pm}$.

When $\lambda_5<0$, one has $M_{T^{\pm\pm}} > M_{H^\pm}
>M_{H^0,A^0}$. In this case,  the decay processes of $T^{--}
\to H^- W^-$ and $H^- \to H^0(A^0)\, W^-$ can be allowed through
the usual gauge interactions;
\newcommand{\lrpartial}
{\overleftarrow{\partial}\hspace{-2.3ex}\overrightarrow{\partial}}
\newcommand{\mE}{E\hspace{-1.3ex}\slash}
\begin{eqnarray}
{\cal L}&=& i g W^+ [ H^+ \lrpartial T^{--}
     + {1\over\sqrt{2}} H^0 \lrpartial H^-
     + {i\over\sqrt{2}} A^0 \lrpartial H^- ]
     \nonumber\\
     && +\; h.c. \,,
\end{eqnarray}
giving rise to the decay rate
\begin{eqnarray}
 \Gamma(T^{--} \to H^- W^- ) &=& {g^2\over 8\pi} M_W
           \left[1+{2y^2-y-1 \over 2} r_W\right] \nonumber\\
  &&         \left[{(y+1)^2\over4} r_W-1 \right]^{1/2}
\end{eqnarray}
where $y\equiv 2|\lambda_5|/g^2$. To suppress this decay mode, we
will require $M_{T^{\pm\pm}}< M_{H^\pm} + M_W$, that is,
$M_{T^{\pm\pm}} > {(y+1)\over2}M_W$.  For $M_{T^{\pm\pm}}= 200$
GeV, it implies $ |\lambda_5|<0.89$. Thus, the decay $T^{--} \to
H^- W^-$ is forbidden unless the coupling $\lambda_5$ is extremely
large.  Now, the off-shell production of $W$, $T^{--} \to H^-
W^{*-}$, is allowed to have the rate;
\begin{equation}
 \Gamma(T^{--} \to H^{-} W^{*-} ) \approx  {3 G_F^2 \over 40 \pi^3}
  {y^5 M_W^{10}\over M_{T^{\pm\pm}}^5}
\end{equation}
in the leading term of $y M_W^2$.  With the further requirement of
$\Gamma(T^{--} \to H^{-} W^{*-} ) < \Gamma(T^{--} \to l_i l_j )$,
we limit ourselves in the parameter space satisfying
\begin{equation}
 |\lambda_5| < 0.16
  \left( M_{T^{\pm\pm}} \over 200 \mbox{ GeV}\right)^{6/5}
        \left( f_{ij} \over 10^{-3} \right)^{2/5} \,.
\end{equation}

Let us now note that the triplet
Higgs decay is short enough to occur inside colliders.
Assuming Eq.~(6) as the main decay rates and
recalling $\sum_{ij} f^2_{ij} \propto \mbox{Tr}(M^2_\nu)$
where $M^\nu_{ij}=f_{ij} \xi v_\Phi$, one obtains the following form
of the total decay rate:
\begin{eqnarray}
&& \Gamma_{T^{\pm\pm}}=  M_{T^{\pm\pm}} \left(
  {1\over 16\pi } {\bar{m}^2 \over  \xi^2 v_\Phi^2} \right.
\nonumber\\
&&~~ \left. + {\alpha_2 \over32} {\xi^2 \over r_W}
              (1-4r_W+12r^2_W) (1-4r_W)^{1/2}  \right)
\end{eqnarray}
where $\bar{m}^2\equiv \sum_i m_i^2 $. When  $M_{T^{\pm\pm}} > 2
M_W$,  one finds the minimum value of the total decay rate given
by
$$ \Gamma_{T^{\pm\pm}}|_{min}=
 {1\over 8\pi } {M_{T^{\pm\pm}}\bar{m}^2 \over \hat{\xi}^2 v_\Phi^2}$$
where $\hat{\xi}^2\equiv  (2\sqrt{2}/g) r^{1/2}_W (\bar{m}/v_\Phi)
(1-4r_W+12 r^2_W)^{-1/2}(1-4r_W)^{-1/4}$. Taking $\bar{m}=0.05$ eV
and $M_{T^{\pm\pm}}=200$ GeV, we obtain
$\Gamma_{\pm\pm}|_{min}\approx 6\times10^{-13}$ GeV and $\hat{\xi}
\approx 6\times10^{-7}$, leading to $\tau|_{max} \approx 0.03$ cm.
When $M_{T^{\pm\pm}} < 2 M_W$, only the first term in Eq.~(12)
contributes and the total decay rate is then $\Gamma >
8\times10^{-14}$ GeV for $ M_{T^{\pm\pm}}=100$ GeV and $\xi <
10^{-6}$. Thus, as far as $T^{--} \to l_i l_j$ are the main decay
modes of the doubly charged Higgs boson, its decay signal should
be observed in colliders.

In the linear collider  with $\sqrt{s}=1$ TeV, the pair production
   cross section is
   $\sigma\approx (100-10)$ fb for $M_{T^{\pm\pm}}
   = (100-450)$ GeV.
   Taking $L=1000$/fb, the number of the produced $T^{\pm\pm}$ will
   be  $N=(10^5-10^4)$.
   In LHC with $L=1000/$fb,  the number of the
   reconstructed pair production  events is expected to be $N=(10^5-10^3)$
   for $M_{T^{\pm\pm}} = (100-450)$ GeV and it becomes down to $N=10$
   for $M_{T^{\pm\pm}} =1000$ GeV.
Thus, both LC and LHC can produce enough numbers of $T^{\pm\pm}$
to probe the neutrino mass pattern if $M_{T^{\pm\pm}} \lsim 450$
GeV. In this case, the precise measurement of the branching ratios
can also reconstruct the neutrino mass matrix $f$. LHC has also a
good potential to confirm the triplet Higgs model as the source of
neutrino mass matrix up to the triplet mass around 1 TeV. Let us
finally note that the observation of the leading decay modes will
be enough to discriminate the neutrino mass patterns: hierarchy
with $m_1<m_2<m_3$ (HI); Inverse Hierarchy with $m_1 \simeq m_2
\gg m_3$ (IH1) and $m_1=-m_2\gg m_3$ (IH2); Degeneracy with $m_1
\simeq m_2 \simeq m_3$ (DG1), $m_1 \simeq m_2 \simeq -m_3$ (DG2),
$m_1 \simeq -m_2 \simeq m_3$ (DG3), $m_1 \simeq -m_2 \simeq -m_3$
(DG4), each of which predicts

$\mbox{(HI)}~~~~~ B(\mu\mu):B(\tau\tau):B(\mu\tau)
={1\over2}:{1\over2}:1$

$\mbox{(IH1)}~~~ B(ee):B(\mu\mu):B(\tau\tau):B(\mu\tau)
=1:{1\over4}$

$\mbox{(IH2)}~~~ B(e\mu):B(e\tau)= 1: 1$

$\mbox{(DG1)}~~ B(ee):B(\mu\mu):B(\tau\tau)=1:1:1$

$\mbox{(DG2)}~~ B(ee):B(\tau\tau)=1:1$

$\mbox{(DG3)}~~ B(e\mu):B(e\tau):B(\mu\tau)=
     1:1:{1\over2}\cot^2\theta_3$

$\mbox{(DG4)}~~ B(\mu\mu):B(\tau\tau):B(e\mu):B(e\tau)=
     {1\over4} \cot^2\theta_3: {1\over4} \cot^2\theta_3:
     1:1$

\section{Supersymmetric Standard Model with R-parity violation}

The general superpotential of the supersymmetric standard model
allowing  lepton number violation  is
$$
W_0 = \mu H_1 H_2 + \bY_e L H_1 E^c + \bY_d Q H_1 D^c
 + \bY_u Q H_2 U^c ,
$$

\begin{equation}  \label{WRpV}
 W_1=  \lambda_{i} L_i L_3 E^c_3 + \lambda'_{i} L_i Q_3 D^c_3,
\end{equation}
where  $W_0$ is R-parity conserving part and $W_1$ is R-parity
violating part written in the basis where the bilinear term $L_i
H_2$ is rotated away.  Here, we have taken only 5 trilinear
couplings, $\lambda_i$ and $\lambda'_i$, assuming the usual
hierarchy of Yukawa couplings. Among soft SUSY breaking terms,
R-parity violating bilinear terms are given by
\begin{equation} \label{VRpV}
 V_0 = m^2_{L_i H_1} L_i H_1^\dagger +  B_i L_i H_2  + h.c.,
\end{equation}
where $B_i$ is the dimension-two soft parameter.  We will denote
the Higgs bilinear term as $B H_1 H_2$.  In the mSUGRA model, the
bilinear parameters, $m^2_{L_i H_1}$ and $B_i$ vanishes at the
supersymmetry breaking mediation scale, and their non-zero values
at the weak scale are generated through renormalization group (RG)
evolution  which will be included in our numerical calculations.
For the consistent calculation of the Higgs and slepton potential,
we need to include the 1-loop contributions to the scalar
potential as follows:
\begin{eqnarray}
V_{1}=\frac{1}{64\pi^2} {\bf Str} {\cal M}^4 \left( \ln\frac{{\cal
M}^2} {Q^2}-\frac{3}{2}\right).
\end{eqnarray}
As is well-known, the electroweak symmetry breaking gives rise to
a nontrivial vacuum expectation values of sneutrino
$\tilde{\nu_i}$ as follows:
\begin{equation}\label{svev}
 \xi_i \equiv \frac{\langle \tilde{\nu_i} \rangle}
  {\langle H_1^0 \rangle} = - {m^2_{L_i H_1} + B_i t_\beta +
\Sigma_{L_i}^{(1)}
          \over m^2_{\tilde{\nu}_i} + \Sigma_{L_i}^{(2)} },
\end{equation}
where the 1-loop contributions $\Sigma_{L_i}^{(1,2)}$ are given by
$\Sigma_{L_i}^{(1)}=\partial V_1/H_1^{\ast}\partial L_i$,
$\Sigma_{L_i}^{(2)}=\partial V_1/L_i^{\ast}\partial L_i$.
The bilinear  R-parity violating parameters induce the mixing
between the ordinary particles and superparticles, namely,
neutrinos/neutralinos,
charged leptons/charginos, neutral Higgs bosons/sneutrinos, as
well as charged Higgs bosons/charged sleptons.
The mixing between neutrinos and neutralinos particularly
serves as the origin of the tree-level neutrino masses. We note
that the parameters $\xi_i$ should be very small to account for
tiny neutrino masses. While the effect of such small parameters on
the particle and sparticle mass spectra (apart from the neutrino
sector) are negligible, they induce small but important R-parity
violating vertices between the particles and sparticles, which
in particular destabilizes the LSP together with the original
trilinear couplings, $\lambda_i$ and $\lambda'_i$.
From the seesaw formulae associated with the heavy four
neutralinos, we obtain the light tree-level neutrino mass matrix
of the form ;
\begin{equation} \label{Mtree}
 M^{tree}_{ij} = -{M_Z^2 \over F_N} \xi_i \xi_j \cos^2\beta ,
\end{equation}
where $F_N= M_1M_2/M_{\tilde{\gamma}}+ M_Z^2 \cos{2\beta}/\mu$ with
$M_{\tilde{\gamma}} = c_W^2 M_1 + s_W^2 M_2$.
The R-parity violating vertices between particles and sparticles
can give rise to 1-loop neutrino masses. Including all the 1-loop
corrections, the loop mass matrix can be written as
 \bea \label{neutrinoloop}
M^{loop}_{ij}= 
-\frac{M_Z^2}{F_N} \left( \xi_i \delta_j +\delta_i \xi_j \right)
\cos\beta +\Pi_{ij},
 \eea
where $\Pi _{ij} $ denotes the 1-loop contribution of the neutrino
self energy and
\bea \label{comentary} \delta_i &=& \Pi_{\nu_i
\widetilde{B}^0} \left(
\frac{-M_2\sin^2\theta_W}{M_{\widetilde{\gamma}} M_W \tan\theta_W}
\right) + \Pi_{\nu_i \widetilde{W}_3} \left( \frac{M_1
\cos^2\theta_W}{M_{\widetilde{\gamma}}M_W} \right) \nonumber \\
&&+\Pi_{\nu_i \widetilde{H}_1^0} \left( \frac{\sin\beta}{\mu}
\right) +\Pi_{\nu_i \widetilde{H}_2^0}\left(
\frac{-\cos\beta}{\mu} \right).
 \eea
Based on the neutrino mass matrix presented in the above, we will
discuss whether the above mass matrix $M^\nu$ can account for both
atmospheric and solar neutrino experimental data.

For the generic parameter space of the R-parity violating mSUGRA
model, the tree mass is dominating well over  the loop
contribution so that the atmospheric neutrino mixing angle
$\theta_{23}$ has much cleaner correlation with the parameter
$\xi_i$ than $\lambda'_i$,  confirming the relation,
$\sin^22\theta_{23} =4 |\xi_2|^2 |\xi_3|^2/\sum_i |\xi_i|^2$. The
condition, $|\lambda'_i| < |\lambda'_2| \approx |\lambda'_3|$, is
required to arrange the large atmospheric angle $\theta_{23}$ and
the small CHOOZ angle $\theta_{13}$. However, the ratio of the
loop and tree masses  is smaller than 0.1 so that it cannot
account for the mild hierarchy between the solar and atmospheric
neutrino mass scales. This implies that there must be some
cancellation to reduce the tree mass.  As a result, the clean
correlation between $\xi_i$ and the angle $\theta_{23}$ is lost,
which makes it difficult to probe neutrino oscillation through the
LSP decays, In fact,  $\lambda'_i$ have the better correlation
than $\xi_i$ for the solution points. Another consequence of the
tree mass suppression is that the loop correction to the sneutrino
VEV  should be taken properly into account for the determination
of the neutrino oscillation parameters in mSUGRA models.

Let us now discuss how the trilinear R-parity violating
couplings are constrained by the neutrino data.
Even though $\lambda'_2/\lambda'_3$ has no
analytic relation with $\theta_{23}$ for the solution points,
we can obtain the following favorable  ranges through
the parameter scan:
$\left|\lambda'_2/\lambda'_3 \right|$ from the neutrino data
\begin{eqnarray}
 0.4 \lsim |\lambda'_2/\lambda'_3| \lsim 2.5 &\mbox{for}&
   \tan\beta = 3-15 \nonumber \\
 0.3 \lsim |\lambda'_2/\lambda'_3| \lsim 3.3 & \mbox{for}&
 \tan\beta =  30- 40.
\end{eqnarray}
It is also amusing to find the correlation between
the ratio $\lambda_2/\lambda_3$
and the solar neutrino mixing angle $\theta_{12}$
Similar to the case of the atmospheric neutrino oscillation, we can get
the constraints:
\begin{eqnarray}
  0.3 \lsim |\lambda_1/\lambda_2| \lsim 1.6 & \mbox{for} &
\tan\beta=3-15, \nonumber \\
 0.2 \lsim |\lambda_1/\lambda_2| \lsim 5.0 & \mbox{for} &
\tan\beta=30-40.
\end{eqnarray}
In addition, fitting the measured mass-squared values,
we find the allowed regions as follows:
\bea
|\lambda_{1,2}|\,, \;\; |\lambda'_{2,3}|
&=& (0.1-2)\times 10^{-4}, \\
|\lambda'_1|  &<& 2.5\times 10^{-5}.
 \eea
The above four equations are the key predictions of the mSUGRA model,
some of
which can be tested in the future colliders.

Since the LSP is destabilized by the R-parity violating interactions,
the structure of the R-parity violating couplings shown above
may be probed by observing the lepton number violating signals
of the LSP decay.  Based on the parameter sets constrained by the
neutrino data,
one can calculate
the cross section for the pair production of the LSP,
which can be either a neutralino or a stau, and
then its decay length and branching ratios.
Taking the luminosity of 1000/fb/yr in the future colliders,
the branching ratios of the order
$10^{-4}-10^{-3}$ will be measurable
as the LSP production cross sections are of
the order 10-100 fb.

{\bf Stau LSP}:
When the LSP is the stau, $\tilde{\tau}_1$, it mainly decays into two leptons
through the coupling $\lambda_i$.
For small $\tan\beta$, $\tilde{\tau}_1$ is almost the right-handed
stau $\tilde{\tau}_R$  due to the small left-right mixing mass.
Then the light stau  almost decays into leptons
via $\lambda_i L_i L_3 E^c_3$ terms in the superpotential.
Thus, one can expect that the branching ratios of those decay
channels depend on the parameter $\lambda_i$.
In this case, the following relation holds,
 \begin{eqnarray}
&& Br(e\nu) : Br(\mu \nu) : Br(\tau \nu)
  \simeq \nonumber\\
&&
|\lambda_1 |^2 : |\lambda_2 |^2:  |\lambda_1 |^2 + |\lambda_2 |^2.
\label{rel1}
 \end{eqnarray}
The corresponding decay length is much
smaller than  micro-meter($\mu m$) so that the stau LSP production
and decay occur instantaneously. The R-parity violating signals in
the linear collider will be
$$e^+ e^- \to \tilde{\tau}_1 \bar{\tilde{\tau}_1}
\to l^+_i l^-_j \nu \bar{\nu},$$  which are identical to the
Standard Model background,
$$e^+ e^- \to W^+ W^- \to l^+_i l^-_j \nu \bar{\nu} \,.$$
But this is a flavor independent process and can be deduced
from the total number of events to establish the flavor dependent
quantities $Br(l_i \nu)$.  Therefore, the observation of the following
relations:
\begin{eqnarray}
 Br(\tau \nu) &=& Br(e\nu) + Br(\mu\nu), \nonumber\\
 {Br(e\nu) \over Br(\mu\nu)} &\approx& 0.1-2.6.
\end{eqnarray}
will be a strong indication of the R-parity violation.

The situation is more complicated for large $\tan\beta$.
A characteristic feature of this case
is that there is a sizable fraction
of the stau LSP decay into top and bottom quarks,
if available kinematically, which is a consequence of
a large left-right stau mixing.  However,
the above relation (\ref{rel1}) becomes obscured by
the large tau Yukawa coupling effect.
The deviation from  (\ref{rel1}) comes from
the large mixing between the stau and the charged Higgs.
Another effect would be the charged Higgs contribution to the event:
$$e^+ e^- \to H^+ H^- \to \tau^+ \tau^- \nu \bar{\nu} \,.$$
Even though a clean prediction for the $\tau$ sector is lost,
we are still able to establish the lepton number violating
signals in the first two generations and measure the quantity:
\begin{equation}
 {Br(e\nu) \over Br(\mu\nu)} =
 \left| {\lambda_1 \over \lambda_2} \right|^2
 \approx 0.04 - 25  \,.
\end{equation}

{\bf Neutralino LSP}:
The lepton number violating signatures from the neutralino decay have
been studied extensively in the literature
as the LSP is a neutralino in the  most
parameter space.
A characteristic feature of the neutralino LSP is that
the vertex for the process $\tilde{\chi}_1^0 \to  l_i W$
is proportional to $\xi_i$ which determines the tree-level neutrino
mass (17).  As a result,
measuring the branching ratios $Br(l_i jj)$ through either on-shell
 or off-shell $W$ bosons will determine the ratio of $|\xi_i|^2$, that is,
\bea Br(ejj)
: Br(\mu jj) : Br( \tau jj) = |\xi_1|^2 : |\xi_2|^2 : |\xi_3|^2.
\eea
If the tree mass dominates over the loop mass, which is usually
the case in the gauge-mediated supersymmetry breaking models,
the neutrino mixing angles $\theta_{23}$ and $\theta_{13}$ can be cleanly
measured in colliders.
Unfortunately, this is not the case in the mSUGRA model under consideration.

As we discussed in the previous section (See Fig.~3), the tree mass
has to be suppressed and thus the variables $\xi_i$ are not correlated with
the mixing angles $\theta_{23}$ and $\theta_{13}$ in general and
$\lambda'_i$ maintain better correlations (See Fig.~6 and 7).
Although $Br(l_i jj)$ or $Br(l_i W)$ are
proportional to $\xi_i^2$, the correlation
with the mixing angles is lost.
On the other hand,  similarly to the stau LSP case,
we can
extract the information on $\lambda_i$ from the measurement of
$\nu l_i^\pm \tau^\mp$ branching ratios for small $\tan \beta$
because the following relation,
\begin{eqnarray}
&&Br(\nu e^{\pm} \tau^{\mp}) :
Br( \nu \mu^{\pm} \tau^{\mp}) : Br(\nu \tau^{\pm} \tau^{\mp})
 \simeq  \nonumber\\
&& |\lambda_1 |^2 : |\lambda_2 |^2:  |\lambda_1 |^2 + |\lambda_2
|^2, \label{ccc}
\end{eqnarray}
holds.
Likewise, if we measure the above branching ratios,
the models can be tested by comparing them with the values allowed
by the neutrino experimental data.  Note that
this is the case for small $\tan\beta < 15$.
For large $\tan\beta$, the relation (21) is invalidated
again because of the large tau Yukawa coupling effect.

Another interesting aspect is that the parameters $\lambda'_i$ can be probed
{\em if the neutralino LSP is heavy enough to allow the decay modes,
$l_i t \bar{b}$}.  The main contributions to these final states
come from the couplings $\lambda'_i$  and thus one obtains the
following approximate relation:
\begin{equation}
 Br(e t\bar{b}) : Br(\mu t\bar{b}) : Br(\tau t\bar{b})  \approx
|\lambda'_1|^2 : |\lambda'_2|^2 : |\lambda'_3|^2
\end{equation}
which should be consistent with the predictions (20), (22) and (23).
One also finds the branching ratios for $l_i t \bar{b}$ large
enough to be measured in the colliders with the integrated luminosity
of $1000/$fb.   The branching ratios get too small if the LSP mass
is more than twice top quark mass.  This is because
the $\nu t\bar{t}$ mode occurring through the light Higgs exchange
becomes dominating.

\section{Supersymmetric Triplet Seesaw Model at High-Energy scale}

Supersymmetric extensions of the Standard Model  exhibit plenty of
new CP violating (CPV) phases in addition to the unique CKM phase.
Although new sources of CP violation are welcome to dynamically
achieve an adequate matter -- antimatter asymmetry, it is known
that they constitute a threat for very sensitive CP tests like
those of the EDMs, at least for supersymmetric masses which are
within the TeV region. In view of such constraints as well as of
those coming from flavor changing neutral current processes, one
can safely assume that all the terms which softly break
supersymmetry are real and flavor universal at the scale where
supersymmetry breaking is communicated to the visible sector. In
spite of this, their renormalization group  running down to the
electroweak scale feels the presence of the Yukawa couplings in
the superpotential which can induce flavor and CP violation in the
soft breaking sector at low energy.

In addition to the logarithmically divergent RG corrections, there
are additional finite contributions to the soft  terms in the
seesaw mechanism.  In type-I seesaw, these contributions are
induced by the bilinear soft-term $\bB_N \bM_N
\tilde{N}\tilde{N}$, associated with the Majorana mass matrix
$\bM_N$  for the heavy singlet states $N$. The superpotential of
the type-I seesaw mechanism reads: $ W =W_0 + W_N$
 with
\be W_N = \bY_N H_2 N L + \frac12 \bM_N NN . \label{mssm} \ee
 Assuming flavor universality and CP conservation
 of the supersymmetric sector,
 finite and infinite  LFV and CPV  radiative
corrections are induced by the new flavor structures ($\bY_N$ and
$\bM_N$). Such contributions  are proportional to the quantity
$\bY^{\dagger}_N \bY_N$. In spite of this dependence on the
``leptonic'' quantity $\bY_N$, it is  worthwhile emphasizing that
these contributions affect also the hadronic EDMs, a point which
was missed in the literature. In particular, this applies to the
finite contributions  to the trilinear
 $\bA_u$ term which is corrected as
\be \label{au-n}
\delta \bA_u = - \frac{1}{16\pi^2} \bY_u
{\rm tr} (\bY^\dagger_N \bB_N \bY_N),
\ee
leading to quark EDMs,
and thus to a nonzero neutron EDM.
The scheme yields the following ratios of EDMs:
\be \label{edmI}
 {d_\mu \over d_e} \approx {m_\mu \over m_e} { (\bY^\dagger_N \bB_N \bY_N)_{22}
  \over (\bY^\dagger_N\bB_N \bY_N)_{11} } ,\quad
  {d_u \over d_e} \approx  C {m_u \over m_e} { {\rm tr}(\bY^\dagger_N \bB_N
\bY_N)
  \over (\bY^\dagger_N  \bB_N \bY_N)_{11} },
\ee
where $C$ is a factor depending on the soft mass parameters.
The above ratios (\ref{edmI}) are strongly
model-dependent given their dependence on the combination
$\bY^\dagger_N \bB_N \bY_N$

A more predictive picture for LFV and CPV can emerge in the
triplet seesaw case. Here  the MSSM superpotential $W_{0}$ is
augmented by
 \be W_T\!=\!\! \frac{1}{\sqrt{2}}(\bY_T L  T L +
\lambda_1 H_1 T H_1 + \lambda_2 H_2\bar T H_2)+ M_T T \bar T
\label{Wt} \ee where the supermultiplets $T\!\!=\!\!(T^0,
T^{+},T^{++})$, $\bar T\!=\!(\bT^{0}, \bT^{-}, \bT^{--})$ are in a
vector-like $SU(2)_W \times U(1)_Y$ representation, $T\sim (3,1)$
and $\bT \sim(3,-1)$. $\bY_T$, a complex symmetric matrix, is
characterized by
   6  independent moduli  and 3 physical phases, while
the parameters $\la_2$
and $M_T$ can be taken to be  real, and  $\la_1$  is in general complex.
After  integrating out the triplet states at the scale $M_T$,
the resulting neutrino
mass matrix becomes
\be
\label{numass}
 M_\nu =\bU^* \bm^D_\nu \bU^\dagger =
 \frac{v^2_2 \la_2}{M_T}\, \bY_T  ,
\ee where $\bm^D_\nu$ is the diagonal neutrino mass matrix and
$\bU$ is the neutrino mixing matrix.

\begin{figure}
\includegraphics[width=0.41\textwidth]{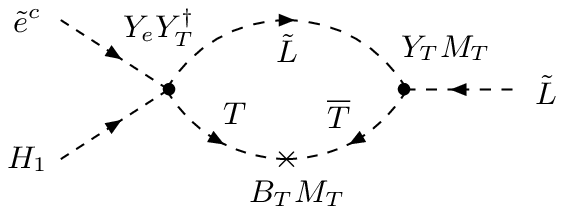}
\caption{\label{fig1} Example of one-loop finite contribution to
the trilinear coupling $\bA_e H_1 \tilde{e}^c \tl$, induced by the
bilinear term $B_T M_T T \bT$.}
\end{figure}

We now turn to the EDM predictions in this model. First of
all, out of the three phases present in the neutrino sector, only the Dirac
phase $\delta$ may entail CP-violating effects
in the LFV entries (this is due to the symmetric nature of $\bY_T$).
However, the contributions to physical observables such as the EDMs
turn out to be quite suppressed in general.
Indeed, due to the hermeticity of  $\bY^\dagger_T \bY_T$,
the phase of the electron EDM amplitude is always proportional to the small
neutrino mixing angle $\theta_{13}$ and to a high power of the Yukawa
couplings. Only in very special circumstances with $\theta_{13}$ close to
the present experimental limit and very large $\tan\beta$, these
contributions could become sizeable.

On the other hand,  a single CP phase residing in the soft term
$B_T M_T T\bar{T}$ can play a significant role in generating
non-zero EDMs, once  we assume  vanishing CP phases in $\mu$ and
tree-level $A$-terms. In such a case, the trilinear couplings
$\bA_e, \bA_d, \bA_u$ receive finite `complex' radiative
corrections at the decoupling of the heavy states $T, \bar{T}$,
exhibiting the common phase from the soft-term $B_T$. In Fig.~1 we
show the  diagrammatic contribution to  $\bA_e$ proportional to
$\bY^\dagger_T \bY_T$. Similar diagrams generate other
contributions proportional to $|\lambda_i|^2$, relevant for
$\bA_e, \bA_d$ and  $\bA_u$. Thus we obtain: \bea
 \delta {\bA}_e & = &
-\frac{3}{16 \pi^2} \bY_e \left(\bY^\dagger_T \bY_T
+ |\la_1|^2\right) B_T , \nonumber\\
\delta {\bA}_d & = &
-\frac{3}{16 \pi^2} \bY_d  |\la_1|^2 B_T,\label{finite} \\
\delta {\bA}_u & = &
-\frac{3}{16 \pi^2} \bY_u  |\la_2|^2 B_T . \nonumber
\eea
The  lepton (quark) EDMs arise from  one-loop diagrams that involve
the exchange of  sleptons (squark) of both chiralities  and Bino (gluino)
(at leading order in  the electroweak breaking effects).
The parametric dependence of the EDMs
at the leading order in the trilinear couplings
goes as follows
\bea
\label{edmsall}
\frac{(d_{e})_i}{e }& \approx&  \frac{-\alpha}{4\pi c^2_W} m_{e_i}
\frac{ M_1{\rm Im} (\delta \hat{ \bA}_{e})_{ii} }{m^4_\tl} F(x_1)
   ,\nonumber\\
\frac{(d_{d})_i}{e}& \approx &  \frac{-2\alpha_s}{9 \pi} m_{d_i}
\frac{M_3 {\rm Im} (\delta \hat{\bA}_d)_{ii}}{m^4_\tq} F(x_3) , \label{edm}\\
\frac{(d_{u})_i}{e}& \approx &  \frac{4 \alpha_s}{9 \pi} m_{u_i}
\frac{M_3{\rm Im} (\delta \hat{\bA}_u)_{ii}}{m^4_\tq} F(x_3) ,
\nonumber \eea where $M_1$ and $M_3$ are the Bino and gluino
masses, respectively, the trilinear couplings have been
parameterized as $\delta {{\bA}}_f = \bY_f \delta \hat{\bA}_f (f=
e, u , d)$, and
 $F(x)$  ($x_1= {M_1^2}/{m^2_\tl},
x_3= {M_3^2}/{m^2_\tq}$) is a loop function of order one. Finally,
by using Eqs.~(\ref{finite},\ref{edmsall}),  we arrive at the
peculiar result, namely the ratio of the leptonic EDMs can be
predicted only in terms of the neutrino parameters: \be
\label{edm-lep1} \frac{d_\mu}{d_e} \!\approx \!\frac{m_\mu}{m_e}
\frac{[\bU (\bm^D_\nu)^2 \bU^\dagger]_{22}} {[\bU (\bm^D_\nu)^2
\bU^\dagger]_{11}} ,\quad \frac{d_\tau}{d_\mu}\!\approx
\!\frac{m_\tau}{m_\mu} \frac{[\bU (\bm^D_\nu)^2 \bU^\dagger]_{33}}
{[\bU (\bm^D_\nu)^2 \bU^\dagger]_{22}}, \ee where $d_e \equiv
(d_e)_1, d_\mu \equiv (d_e)_2$ {\it etc}, and for simplicity we
have assumed $|\la_1|^2 \ll (\bY^\dagger_T \bY_T)_{ii}$.
Notice that the presence of extra CPV phases would alter the
simple form of the above ratios (\ref{edm-lep1}) and in general
the result would be more model dependent. Regarding some numerical
insight, we can consider three different neutrino mass patterns as
before.  For each case, the relative size of the entries in
Eq.~(\ref{edm-lep1}) is given as follows: \bea
 {[\bV (\bm^D)^2_\nu \bV^\dagger]_{11}}:
{[\bV (\bm^D)^2_\nu \bV^\dagger]_{22}}:
{[\bV (\bm^D)^2_\nu \bV^\dagger]_{33}}\nonumber \\
 = \cases{  c_{13}^2 s_{12}^2 + \rho s_{13}^2  : \rho c_{13}^2 s_{23}^2
        : \rho c_{13}^2 c_{23}^2  & \mbox{(HI)} \cr
        c_{13}^2 : c_{23}^2 : s_{23}^2 & \mbox{(IH)} \cr
        1 : 1 : 1 & \mbox{(DG}) \cr } \label{YdagY}
\eea where $\rho= \!\frac{ \Delta m^2_{31}}{\Delta m^2_{21}}\sim
25$
 and
$s_{ij}\, (c_{ij}) \!= \!\sin\theta_{ij} \, (\cos\theta_{ij})$.
Therefore, according to Eq.~(\ref{edm-lep1}) and using the present
best fit neutrino parameters  with
 $s_{13}\ll 0.1$,
we obtain the following leptonic  EDM ratios:
\bea
 \frac{d_\mu}{d_e} \approx \frac{m_\mu}{m_e}\frac{\rho s^2_{23}
}{s_{12}^2}
 \sim 10^4 , && ~
\frac{d_\tau}{d_\mu} \approx \frac{m_\tau}{m_\mu}
\frac{ s^2_{23}}{c^2_{23}}
\sim 17 ,   ~~({\rm HI}) \nonumber \\
\!\!\!\phantom{dd} \frac{d_\mu}{d_e} \approx \phantom{\rho} \frac{m_\mu}{m_e} c^2_{23}
 \sim 10^2 , && ~ 
\frac{d_\tau}{d_\mu} \approx \frac{m_\tau}{m_\mu}
\frac{ s^2_{23}
}{c^2_{23}}
\sim 17 ,   ~~({\rm IH}) \nonumber   \\
\!\!\!\phantom{dd}\frac{d_\mu}{d_e} \approx \frac{m_\mu}{m_e}
 \sim 2\times 10^2 , && ~
\!\!\!\! \frac{d_\tau}{d_\mu} \approx \frac{m_\tau}{m_\mu} \sim 17
,   ~~({ \rm DG}) .  \label{edmHI}  \eea

We are now tempted to  give an order-of-magnitude estimate
of $d_e$ to show that sizeable
values can be attained:
\be
\frac{d_e}{e} \sim  10^{-29}
\left(\frac{M_T}{10^{11}~{\rm GeV}}\cdot
\frac{10^{-4}}{\lambda_2}\right)^2
\left ( \frac{200~{\rm GeV}}{\tm}\right)^{2}~
{\rm cm}
\ee
where  we have taken a common SUSY mass scale,
$M_1 =  m_\tl ={\rm Im}(B_T)= \tm $ and the pattern HI.
This shows that the electron and muon EDMs could be within
the future experiment reach (see Table~1).
We also notice from eq.~(\ref{finite})
that lepton and quark EDMs are definitely correlated
in this scenario. However,  such a correlation is also sensitive
to the ratio $M_T/\lambda_2$ and to
other mass parameters, such as the gaugino,  squark and slepton
masses, and so can only be established in a specific SUSY breaking
framework.

Another interesting prediction regards the relative size of LFV
among different flavors. For instance, the ratio of the LFV
entries of the left-handed slepton mass matrix is: \be
\label{predi}
 \frac{ (\bm^{2 }_{\tilde{L}})_{\tau \mu}}
  {(\bm^{2 }_{\tilde{L}})_{\mu e} } \approx
\frac{ (\bY^{\dagger}_T \bY_T)_{23} }
  { (\bY^{\dagger}_T \bY_T)_{12} } \approx
\rho
\frac{\sin 2\theta_{23}}{\sin 2\theta_{12}
\cos\theta_{23}} \sim 40
\ee
which holds for  $s_{13} \ll \rho^{-1} c_{12} s_{12} \sim 0.02 $.
This implies that also the branching ratios
$B(\ell_i \to \ell_j  \gamma)$
can be related in terms of only the low-energy neutrino parameters
and  we find
\be \label{brs} {\rm B}(\mu \to e \gamma) : {\rm
B}(\tau \to e \gamma) : {\rm B}(\tau \to \mu \gamma) \sim 1 :
10^{-1} : 300 . \ee
This result does not depend on the detail of
the model, such as either $M_T$ or the SUSY spectrum. On the
contrary, the individual branching ratios in (\ref{brs}) also
depend on quantities such as $\mu, \tan\beta$ and soft SUSY
parameters, which are not of direct concern in the present
discussion of the EDMs.

\vspace{2ex}
\begin{center}
\begin{tabular}{ccc}
\hline
EDM & Present limits & Future limits \\[0.2pt]
\hline
$d_e$ &  $7\times 10^{-28}$ &  $10^{-32}$ \\
$d_\mu$ & $3.7\times 10^{-19}$ &
$10^{-24} - 5\times 10^{-26}$ \\
${\rm Re}(d_\tau)$  & $4.5\times 10^{-17}$ & $10^{-17} - 10^{-18}$
\\ \hline
$d_n$&  $6\times 10^{-26}$ & ?? \\
\hline
\end{tabular}

\vspace{1ex} { TABLE II: Present bounds and future sensitivity (in
{\rm e$\cdot$cm} units) on lepton and neutron EDMs. } \label{tb3}
\end{center}

The presence of triplet states with mass smaller than
the grand unification scale $M_G$
precludes the gauge-coupling unification.
This can be recovered, for instance,
by completing a GUT representation
where $T, \bar{T}$  fit.
 In such a case care should be taken in evaluating the
radiative effects as the additional components of the full GUT multiplet
where triplets reside would also contribute to LFV as well as CPV processes.





\end{document}